\documentclass[11pt]{article}  % Single-column, standard arXiv format

\usepackage{times}  % Use Times New Roman font
\usepackage{natbib}
\usepackage{url}    % Handle URLs properly
\usepackage[utf8]{inputenc} % UTF-8 encoding
\usepackage[small]{caption} % Small captions for figures/tables
\usepackage{graphicx} % Include images
\usepackage{amsmath, amssymb, amsthm} % Math packages
\usepackage{booktabs} % Better tables
\usepackage{algorithm} % Algorithms
\usepackage{algorithmic} % Algorithmic environment
\usepackage{subcaption} % Subfigures
\usepackage{multirow} % Multirow tables
\usepackage[margin=1in]{geometry} % 1-inch margins as per arXiv guidelines

\title{Towards Hybrid Traffic Laws for Mixed Flow of Human-Driven Vehicles and Connected Autonomous Vehicles}

\author{
    Tal Kraicer\\
    Technion - Israel Institute of Technology\\
    \texttt{talkraicer@campus.technion.ac.il} \\
    \and
    Jack Haddad\\
    Technion - Israel Institute of Technology\\
    \texttt{jh@technion.ac.il} \\
    \and
    Erez Karpas\\
    Technion - Israel Institute of Technology\\
    \texttt{karpase@technion.ac.il} \\
    \and
    Moshe Tennenholtz\\
    Technion - Israel Institute of Technology\\
    \texttt{moshet@technion.ac.il}
}

\begin{document}
\maketitle

\begin{abstract}
Hybrid traffic laws represent an innovative approach to managing mixed environments of connected autonomous vehicles (CAVs) and human-driven vehicles  (HDVs) by introducing separate sets of regulations for each vehicle type. These laws are designed to leverage the unique capabilities of CAVs while ensuring both types of cars coexist effectively, ultimately aiming to enhance overall social welfare. This study uses the SUMO simulation platform to explore hybrid traffic laws in a restricted lane scenario. It evaluates static and dynamic lane access policies under varying traffic demands and CAV proportions. The policies aim to minimize average passenger delay and encourage the incorporation of autonomous vehicles with higher occupancy rates. Results demonstrate that dynamic policies significantly improve traffic flow, especially at low CAV proportions, compared to traditional dedicated bus lane strategies. These findings highlight the potential of hybrid traffic laws to enhance traffic efficiency and accelerate the transition to autonomous technology.
\end{abstract}

\textbf{Keywords}: Automated and Connected Driving, Hybrid Traffic Laws, Mixed Traffic Environment, Simulation of Urban MObility (SUMO), Traffic Flow Optimization.

\section{Introduction}
\label{sec:intro}
Existing traffic regulations are fundamentally designed to govern the behavior of human-driven vehicles (HDVs). These regulations are built upon the foundation of human perception, decision-making, and reaction times, often incorporating safety margins to account for human error and variability in driving styles \citep{intro1,intro2}. Furthermore, they are usually created assuming the worst-case scenario driver behavior. For example, speed limits are determined by considering factors such as visibility ranges, road curvature, and the time required for a human driver to perceive and react to unexpected obstacles. These limits are calibrated to human visual acuity and the cognitive processing speed of an average driver \citep{intro3}.

Autonomous vehicles (CAVs), or self-driving cars, are transforming transportation \citep{intro4}. CAVs have abilities beyond those of human drivers, such as strictly following traffic laws, processing data rapidly, and communicating with other vehicles quickly \citep{intro5}. Although these capabilities are still developing, they hold much promise. While CAVs may eventually outperform humans, this work focuses on the near future, assuming CAVs can travel as well as, but not better than, competent human drivers. This realistic approach allows for a more solid analysis of policy impacts in the short term. Additionally, we assume that the integration of CAVs into the infrastructure will be a gradual process. Consequently, we analyze the impact of the proposed policies across a range of CAV proportions.

We adopt a multi-agent system \citep{mas_book} view of transportation, treating each vehicle as its own agent. This allows us to think of traffic laws as a special case of social laws \citep{DBLP:conf/ijcai/TennenholtzM89,DBLP:conf/aaai/ShohamT92}. From this perspective, we argue that the unique capabilities of connected autonomous vehicles (CAVs) allow us to develop a new set of traffic laws tailored to a mixed-traffic environment of both CAVs and human-driven cars. 

Our proposed {\em hybrid} traffic laws differentiate between CAVs and human-driven vehicles and are thus relevant for mixed traffic. By recognizing advanced CAV capabilities, hybrid laws can be designed to optimize traffic flow, enhance safety, and improve overall social welfare. Unlike methods that rely on better driving abilities of CAVs, our hybrid traffic laws rely on the ability of CAVs to obey complex rules that may change dynamically. Differently from other works \citep{intro_CAVS_1},\citep{intro_CAVS_2}, the hybrid traffic laws presented in this paper are more strategic than tactical or operational (as defined in \citep{intro_AV_levels}). Strategic decisions involve long-term planning and overarching goals, tactical decisions focus on medium-term actions and coordination, and operational decisions handle real-time, immediate responses.

This work focuses on a case study of a roadway containing lane reduction and experiencing high traffic demand, resulting in congestion, delays, and reduced overall efficiency. Current traffic management strategies to mitigate such issues often involve the implementation of restricted lanes, such as Dedicated Bus Lanes (DBLs) and High-Occupancy Lanes (HOLs). These approaches aim to minimize average passenger delay by prioritizing the movement of vehicles carrying a greater number of occupants (for a formal definition, see Equation \ref{eq:APD}). Furthermore, these dedicated lanes incentivize the use of public transportation and encourage carpooling, contributing to a reduction in the overall number of vehicles on the road \citep{intro6}.

However, in many real-world scenarios, the effectiveness of DBLs is hampered by an insufficient number of public transport vehicles, leading to underutilization of the designated lane (see data on \citep{tiltan_ayalon}). Similarly, while HOLs have demonstrated their potential to improve traffic flow (see results in Section~\ref{sec:results}), they are unenforceable, as human drivers may violate the occupancy requirements with limited risk of detection \citep{intro8,intro9}.

To address these limitations, this paper proposes a novel hybrid traffic law that leverages the unique capabilities of connected autonomous vehicles (CAVs). Under this proposed regulation, only public transport vehicles and CAVs carrying a minimum number of passengers would be permitted to utilize the restricted lane. This approach capitalizes on the inherent ability of CAVs to adhere strictly to traffic regulations, thus ensuring enforceability. A key assumption underlying this approach, which will be further elaborated upon in the methodology section, is the ability to reliably differentiate between CAVs and human-driven vehicles (HDVs) in real-time traffic conditions.

\label{goals}
The primary goals of implementing this hybrid traffic law are twofold:
\begin{enumerate}
    \item Improve current traffic flow, as measured by a reduction in average passenger delay (APD).
    \item Incentivize a shift towards the use of CAVs and promote carpooling. 
\end{enumerate}
By granting lane access exclusively to CAVs with higher occupancy, this policy creates a clear benefit for those who choose to travel in such vehicles, potentially accelerating the incorporation of autonomous technology and encouraging more efficient travel patterns.

\section{Methodology}
\label{sec:methodology}
This work introduces a hybrid traffic regulation strategy that designates a lane segment exclusively for public transport and connected autonomous vehicles with a minimum passenger count. Positioned before a lane reduction—commonly a congestion bottleneck—this restricted section aims to prioritize high-passenger-count vehicles, thereby reducing the average passenger delay (APD) as defined in Equation~\ref{eq:APD}. This study isolates the policy's direct effects by focusing on a simplified road network while minimizing real-world complexities and noise.

Restricting access to the restricted lane creates a bottleneck at the point where the lane is introduced. However, congestion at the beginning of the restriction point is reduced compared to the original scenario, as only a portion of vehicles need to switch lanes. Merging at the merging point becomes smoother since only a small fraction of vehicles use the restricted lane. Consequently, the restriction allows more efficient road utilization. The results presented in Section~\ref{sec:results} will demonstrate the effectiveness of this approach in improving traffic flow and reducing delays, validating the claims made here.

\begin{figure*}[t]
    \centering
        \centering
        \includegraphics[width=\linewidth]{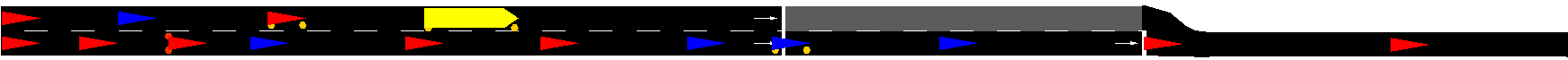}
        \label{fig:network_overview}

    \caption{Illustration of the Configuration of the Road Network.}
    \label{fig:road_network}
\end{figure*}

The road network illustrated in Figure~\ref{fig:road_network} depicts a generalized lane reduction scenario. Initially, the road consisted of two general-purpose (GP) lanes extending for 1 kilometer. About 500 meters before the merge point, the left lane transitions into a restricted lane, permitting access to varying vehicle populations, before ultimately merging into a single lane. The speed limit on the whole road is 25 m/s. The experiments in this work were conducted using the Simulation of Urban MObility (SUMO) platform.

\begin{figure}[t]
    \centering
    \includegraphics[width=0.5\linewidth]{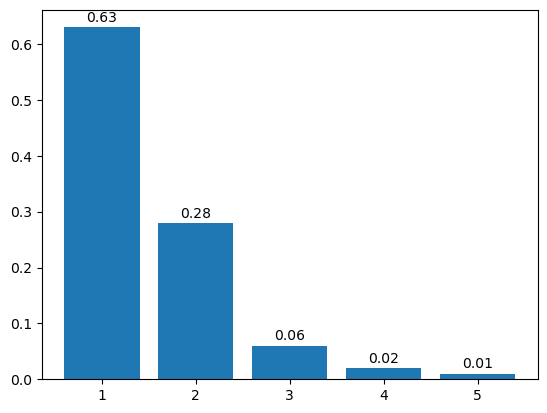}
    \caption{Distribution of the Number of Passengers in a Car, Based on Real Data from NHTS.}
    \label{fig:PassDist}
\end{figure}

Various demand scenarios were tested on the network. The demand parameters are defined as follows:
\begin{itemize}
    \item \textbf{Hourly Vehicle Count (\( \lambda_v \)):} The average number of vehicles entering the network per hour. Two sets of hourly vehicle counts were tested:
    \begin{itemize}
        \item \textit{Daily\_k} \label{def:Daily}: Derived from data on Ayalon North Road in Tel-Aviv at measuring point 2 (defined in \citep{tiltan_ayalon}) in June 2022. The data depicted in Figure \ref{fig:DemandFig} spreads over 14 hours (6:00-20:00) and contains a morning peak and a smaller evening peak. The amounts were scaled down by a factor $k \in \{2,2.5,3\}$ to fit the two-lane configuration, instead of the original five or four lanes in Ayalon road. These demand profiles enable us to simulate a realistic scenario in which traffic demand varies over time, reflecting the behavior of real road users
        \item \textit{Constant\_3000}: A simplified case with constant expected vehicle counts of 3000 per hour over a 1-hour interval. This medium-high demand profile enables us to analyze the behavior of our policy and its impacts, rather than attributing the outcomes to the dynamic nature of the demand profile presented in \textit{Daily\_k}.
    \end{itemize}
    \item \textbf{Vehicle Type Distribution:} The probability distribution of vehicle types:
    \label{def:vTypes}
    \begin{itemize}
        \item \( P(\text{HDV})\): Proportion of Human-Driven vehicles (appear in Figure \ref{fig:road_network} highlighted in red).
        \item \( P(\text{CAV})\): Proportion of Autonomous Vehicles (CAVs) - referred to as the \textit{proportion} (appear in Figure~\ref{fig:road_network} highlighted in blue).
        \item \( P(\text{Bus})\): Proportion of buses (appear in Figure~\ref{fig:road_network} highlighted in yellow).
    \end{itemize}
    \item \textbf{Arrival Process:} To obtain statistically significant results, we generated random data by assuming that vehicle arrivals follow a Poisson process, governed by the hourly demand data mentioned above. The number of vehicles arriving during each interval follows a Poisson process with parameter \( \lambda_v(t) \).
    The time between two consecutive arrivals is exponentially distributed in such an arrival process, $\sim Exp(\lambda_v(t))$.
    \item \textbf{Passenger Count Distribution:}
    \begin{itemize}
        \item \textbf{Cars:} The distribution of the number of passengers in cars is depicted in Figure \ref{fig:PassDist}, based on real data from \citep{nhts_trips}.
        \item \textbf{Buses:} The average number of passengers per bus is 7.05, based on data from \citep{tiltan_ayalon}. Since buses always have access to the restricted lane, only the expected value is considered, without requiring distribution details.
    \end{itemize}
\end{itemize}

All experiments use the same random seed across different policies to ensure an identical arrival process for better comparison. Each experiment is repeated ten times to provide statistical reliability.

To determine which vehicles are permitted to use the restricted lane presented in Figure \ref{fig:road_network}, a set of policies was tested as part of developing hybrid traffic laws:
\begin{itemize}
    \item DBL: A standard Dedicated Bus Lane (DBL), where only buses can use the restricted lane.
    \item Plus\_$i$:  A set of policies where $i \in {1,2,\dots,5}$ represents the minimum number of passengers a vehicle must have to access the lane. Under this rule, any vehicle meeting the threshold can use the lane, making enforcement infeasible for the HDV vehicle population.
    \item CAVStaticPlus\_{i}: A set of policies where $i \in {1,2,\dots,5}$ represents the minimum number of passengers required. This rule allows only CAVs that meet the threshold to enter the lane, ensuring enforceability.
    \item CAVDynamic: Dynamic policies that adjust the minimum passenger threshold required for a vehicle to access the restricted lane based on real-time speed measurements within the restricted lane. These policies operate with a specific speed parameter: if the lane speed falls below the speed parameter, the threshold increases (allowing fewer vehicles), and if it exceeds the speed parameter, the threshold decreases. A threshold adjustment is made every 60 seconds (empirically chosen). For example, \textit{CAVDynamic\_22} uses a speed parameter of 22 m/s, increasing the threshold when speeds drop below 22 m/s and decreasing it when speeds exceed 22 m/s. While other dynamic policies leveraging different features were tested, they demonstrated poor performance compared to this approach.
\end{itemize}

The primary goal of these policies is to minimize the Average Passenger Delay (APD). It is defined based on the following parameters:

\begin{itemize}
    \item \( D_a \): Actual travel time, i.e., the total time a vehicle spends in the system (seconds).
    \item \( D_t \): Free flow travel time (seconds).
    \item \( \text{timeLoss} = D_a - D_t \).
    \item \( \text{departDelay} \): Time a vehicle waits before entering the road due to road capacity limitations (seconds).
    \item \( \text{Vehicle Delay (VD)} = \text{timeLoss} + \text{departDelay} \).
    \item \( \#_{\text{vehicle}} \): Number of passengers in a vehicle.
\end{itemize}

The \textbf{Average Passenger Delay (APD)} is given by:

\begin{equation}
\label{eq:APD}
\text{APD} = \frac{\sum_{\text{all vehicles}}{\text{VD} \times \#_{\text{vehicle}}}}{\sum_{\text{all vehicles}}{\#_{\text{vehicle}}}}
\end{equation}

\label{def:Robust}
The APD metric (passenger delay divided by total number of passengers) focuses on reducing the delay experienced by the average traveler, accounting for both time spent in the system and time waiting to enter it. It is important to consider the time waiting to enter the system, as different policies may cause congestion near the entrance to the road, which will affect the departDelay. Our goal is to develop a \textit{robust} policy that enhances the APD across all demand profiles and vehicle type distributions. This approach is necessary because adjusting the policy for every new scenario is challenging in real-world settings.

To implement this system in a real-world environment, several key assumptions must be considered:
\begin{enumerate}
\item Measuring state variables: The ability to measure the defined state variables (e.g., speed) is essential. This can be achieved using sensors or through direct data reporting from CAVs to a central server.
\item Differentiating between CAVs and HDVs: It is assumed that advanced smart cameras or other identification technologies are available to distinguish between CAVs and HDVs, for example, using vehicle license plate recognition systems. This allows the enforceability of the restricted lane.
\item CAV communication for threshold updates: CAVs are assumed to have the capability to communicate with a central server, enabling them to receive real-time updates on threshold values and adjust their behavior accordingly.
\end{enumerate}

\section{Results and Discussion}

\label{sec:results}
\subsection{Enhancing Network Throughput}
\begin{table*}[ht!]
\centering
    % subfigures for Daily_3:
    \begin{subtable}{\textwidth}
    \centering
    \resizebox{0.7\textwidth}{!}{%
    \begin{tabular}{lcccccc}
    & DBL & Plus\_1 & Plus\_2* & Plus\_3* & Plus\_4* & Plus\_5* \\
    \hline
    \textbf{APD} & \underline{622.23} & 1284.12 & 665.18 & \textit{74.61} & 301.78 & 523.53 \\
    \end{tabular}%
    }
    \caption{Baseline Policies (only HDVs) - \textit{Daily\_3}}
    \label{tab12:Daily_3_baseline}
    \end{subtable}

    % \hspace{0.03\textwidth} % Adjust spacing between tables
    \begin{subtable}{\textwidth}
      \centering
      \resizebox{\textwidth}{!}{%
\begin{tabular}{l|cccccccccc}
    \textbf{Policy / CAVs Proportion} & \textbf{10\%} & \textbf{20\%} & \textbf{30\%} & \textbf{40\%} & \textbf{50\%} & \textbf{60\%} & \textbf{70\%} & \textbf{80\%} & \textbf{90\%} & \textbf{100\%} \\
    \hline
    CAVStaticPlus\_1 & 64.40 & 96.93 & 417.75 & \textbf{814.21} & \textbf{1138.55} & \textbf{1214.03} & \textbf{1200.22} & \textbf{1234.15} & \textbf{1308.01} & \textbf{1284.12} \\
    CAVStaticPlus\_2 & 272.70 & 105.55 & 55.78 & 51.19 & 73.94 & 139.95 & 254.57 & 398.74 & 520.68 & \textbf{665.18} \\
    CAVStaticPlus\_3 & 522.35 & 411.34 & 329.84 & 277.80 & 224.15 & 156.98 & 130.69 & 107.47 & 90.82 & 74.61 \\
    CAVStaticPlus\_4 & 601.41 & 549.33 & 520.78 & 515.31 & 430.67 & 410.81 & 413.61 & 392.67 & 329.91 & 301.78 \\
    CAVStaticPlus\_5 & 606.35 & 596.45 & 599.69 & 593.69 & 556.71 & 549.08 & 525.56 & 525.02 & 521.62 & 523.53 \\
    CAVDynamic\_22 & 74.37 & 100.31 & 329.75 & 435.17 & 472.04 & 372.93 & 502.79 & 467.72 & 515.38 & 524.88 \\
    CAVDynamic\_23 & 80.97 & 102.35 & 242.19 & 317.33 & 369.81 & 404.20 & 423.84 & 472.52 & 450.01 & 447.10 \\
    CAVDynamic\_24 & 72.71 & 102.26 & 204.44 & 253.51 & 266.03 & 264.72 & 293.06 & 326.80 & 336.15 & 316.33 \\
    CAVDynamic\_25 & 74.07 & 94.97 & 159.70 & 196.04 & 218.68 & 208.93 & 243.70 & 250.20 & 272.67 & 298.02 \\

\end{tabular}
% \vspace{2ex}
%
}
    \caption{Dynamic and Static Policies  - \textit{Daily\_3}}
    \label{tab12:daily_3_dynamic}
    \end{subtable}
    \vspace{1ex}

    % Subfigures for Daily_2.5:
    \begin{subtable}{\textwidth}
    \centering
    \resizebox{0.7\textwidth}{!}{%
    \begin{tabular}{lcccccc}
    & DBL & Plus\_1 & Plus\_2* & Plus\_3* & Plus\_4* & Plus\_5* \\
    \hline
    \textbf{APD} & \underline{5237.98} & 6299.91 & 5622.25 & \textit{2882.67} & 4348.99 & 4879.44 \\
    \end{tabular}%
    }
    \caption{Baseline Policies (only HDVs) - \textit{Daily\_2.5}}
    \label{tab12:daily_2.5_baseline}
    \end{subtable}
    % \hspace{0.03\textwidth} % Adjust spacing between tables
    
    \begin{subtable}{\textwidth}
    \centering
    \resizebox{\textwidth}{!}{%
   \begin{tabular}{l|cccccccccc}
    \textbf{Policy / CAVs Proportion} & \textbf{10\%} & \textbf{20\%} & \textbf{30\%} & \textbf{40\%} & \textbf{50\%} & \textbf{60\%} & \textbf{70\%} & \textbf{80\%} & \textbf{90\%} & \textbf{100\%} \\
    \hline
    CAVStaticPlus\_1 & 2775.80 & 3305.57 & 4766.20 & \textbf{5881.69} & \textbf{6316.03} & \textbf{6398.49} & \textbf{6327.12} & \textbf{6358.09} & \textbf{6368.99} & \textbf{6299.91} \\
    CAVStaticPlus\_2 & 4155.44 & 3199.45 & 2651.84 & 2631.48 & 3052.58 & 3636.64 & 4201.79 & 4721.06 & 5214.02 & \textbf{5622.25} \\
    CAVStaticPlus\_3 & 4984.54 & 4696.03 & 4436.90 & 4194.44 & 3988.03 & 3705.87 & 3490.59 & 3291.36 & 3093.36 & 2882.67 \\
    CAVStaticPlus\_4 & 5148.71 & 5104.23 & 4987.73 & 4854.40 & 4780.42 & 4694.15 & 4593.79 & 4519.39 & 4435.66 & 4348.99 \\
    CAVStaticPlus\_5 & 5196.95 & 5120.90 & 5152.12 & 5068.39 & 5058.09 & 5102.86 & 5027.29 & 5006.12 & 4987.16 & 4879.44 \\
    CAVDynamic\_22 & 2829.07 & 3307.65 & 4325.90 & 4626.76 & 4714.05 & 4818.51 & 4741.75 & 4697.69 & 4869.03 & 4885.22 \\
CAVDynamic\_23 & 2861.25 & 3227.78 & 4144.80 & 4362.01 & 4609.12 & 4574.87 & 4551.78 & 4465.73 & 4634.06 & 4828.97 \\
CAVDynamic\_24 & 2861.81 & 3291.28 & 3934.34 & 4232.49 & 4302.61 & 4331.80 & 4326.16 & 4418.85 & 4445.40 & 4443.11 \\
CAVDynamic\_25 & 2977.70 & 3288.84 & 3843.33 & 4115.29 & 4124.72 & 4104.62 & 4108.70 & 4168.42 & 4284.76 & 4188.85 \\

\end{tabular}%
% \vspace{2ex}

    }
    \caption{Dynamic and Static Policies - \textit{Daily\_2.5}}
    \label{tab12:daily_2.5_dynamic}

    \end{subtable}
    \vspace{1ex} % Adjust vertical spacing

    % subfigures for Daily_2:
    \begin{subtable}{\textwidth}
    \centering
    \resizebox{0.7\textwidth}{!}{%
    \begin{tabular}{lcccccc}
    & DBL & Plus\_1 & Plus\_2* & Plus\_3* & Plus\_4* & Plus\_5* \\
    \hline
    \textbf{APD} & \underline{12509.23} & 13907.27 & 13400.91 & \textit{9728.62} & 11440.02 & 12171.34 \\
    \end{tabular}%
    }
    \caption{Baseline Policies (only HDVs) - \textit{Daily\_2}}
    \label{tab12:daily_2_baseline}
    \end{subtable}
    % \hspace{0.03\textwidth} % Adjust spacing between tables

    \begin{subtable}{\textwidth}
      \centering
      \resizebox{\textwidth}{!}{%
        \begin{tabular}{l|ccccccccccc}
            \textbf{Policy / CAVs Proportion} & \textbf{10\%} & \textbf{20\%} & \textbf{30\%} & \textbf{40\%} & \textbf{50\%} & \textbf{60\%} & \textbf{70\%} & \textbf{80\%} & \textbf{90\%} & \textbf{100\%} \\
            \hline
        CAVStaticPlus\_1 & 9570.94 & 10434.51 & 12535.66 & \textbf{13742.24} & \textbf{14180.84} & \textbf{14158.71} & \textbf{14010.83} & \textbf{13921.37} & \textbf{13965.78} & \textbf{13907.27} \\
        CAVStaticPlus\_2 & 11223.45 & 10092.46 & 9472.68 & 9450.08 & 10115.75 & 10903.54 & 11664.73 & 12301.68 & \textbf{12914.75} & \textbf{13400.91} \\
        CAVStaticPlus\_3 & 12156.70 & 11873.63 & 11579.42 & 11285.50 & 10979.13 & 10681.01 & 10416.42 & 10175.09 & 9944.67 & 9728.62 \\
        CAVStaticPlus\_4 & 12399.15 & 12300.18 & 12163.56 & 12052.19 & 12044.36 & 11874.59 & 11784.37 & 11691.48 & 11574.32 & 11440.02 \\
        CAVStaticPlus\_5 & 12479.14 & 12450.63 & 12406.30 & 12389.03 & 12427.17 & 12301.07 & 12248.17 & 12223.47 & 12158.86 & 12171.34 \\
        CAVDynamic\_22 & 9708.91 & 10280.21 & 11812.24 & 12447.89 & 12326.51 & 12462.63 & 12377.29 & 12361.78 & 12476.93 & 12405.53 \\
CAVDynamic\_23 & 9717.59 & 10321.33 & 11479.91 & 12136.50 & 12098.51 & 11986.29 & 12286.10 & 12045.64 & 12232.87 & 12256.05 \\
CAVDynamic\_24 & 9705.37 & 10226.77 & 11296.26 & 11732.61 & 11721.49 & 11727.96 & 11781.21 & 11868.54 & 11910.43 & 11901.82 \\
CAVDynamic\_25 & 9775.91 & 10257.83 & 11022.40 & 11520.51 & 11474.68 & 11582.24 & 11510.73 & 11571.32 & 11586.74 & 11653.17 \\

        \end{tabular}%
      }
    \caption{Dynamic and Static Policies  - \textit{Daily\_2}}
      \label{tab12:daily_2_dynamic}
    \end{subtable}
    \footnotesize{* represents unenforceable policies.}
      \caption{
      APD evaluated across various CAV proportions under the \textit{Daily\_k} demand scenario. 
      The results highlight the best policies: \underline{underline} for the best enforceable policy without CAVs, \textit{italic} for the best unenforceable policy without CAVs, and \textbf{bold} for enforceable policies with CAVs that are outperformed by \underline{underline}.
    }
    \label{tab12:APD}
% \vspace{6ex}

\end{table*}

Table \ref{tab12:APD} shows the APD for the \textit{Daily} demand profiles outlined in Section~\ref{def:Daily}, calculated as the average over 10 simulation runs. The table includes all baseline policies designed exclusively for HDV vehicles. Some of these policies are unenforceable, as discussed in Section~\ref{sec:intro}. The DBL is identified as the most effective among the enforceable HDV policies, while the optimal unenforceable policy is Plus\_3.

As shown in \ref{tab12:daily_3_dynamic}, \ref{tab12:daily_2.5_dynamic} and \ref{tab12:daily_2_dynamic}, simple policies like \textit{CAVStaticPlus\_3} consistently outperform the DBL baseline across all tested scenarios. Given their simplicity and ease of implementation under the assumptions of this study, static policies present a promising solution for immediate incorporation in the near future.

The entries highlighted in bold in Table \ref{tab12:daily_3_dynamic} represent instances where static enforceable CAV policies are outperformed at certain proportions by the best enforceable HDV policy, DBL. Examining these entries demonstrates the limitations of static CAV policies under specific conditions. However, our dynamic CAV policy consistently outperforms DBL and, in some cases, even surpasses the performance of the optimal unenforceable policy, Plus\_3. Across all scenarios in Table \ref{tab12:APD}, the dynamic policies demonstrate consistent improvements over the DBL baseline for every demand profile and proportion, highlighting their robustness (as defined in Section ~\ref{def:Robust}). Moreover, our dynamic policy is robust to the exact choice of control parameters and consistently outperforms the baseline policies across various parameter configurations.

It is worth noting that the performance of the proposed dynamic policies diminishes as CAV proportions increase. In Tables \ref{tab12:daily_2.5_dynamic} and \ref{tab12:daily_2_dynamic}, representing scenarios with higher demand, certain static policies not only outperform DBL but also exceed the performance of dynamic policies at high CAV proportions. This phenomenon can occur because the dynamic policy is limited to a small set of actions. As detailed in Section \ref{sec:methodology}, the policy is restricted to modifying the threshold by only one unit every 60 seconds. When considered alongside the light-tailed occupancy distribution depicted in Figure \ref{fig:PassDist}, such single-step changes can lead to disproportionately large effects, particularly in scenarios involving numerous CAVs. For instance, decreasing the threshold from 2 to 1 enlarges the proportion of CAVs permitted on the restricted lane from 37\% to 100\%, a change that may result in severe traffic congestion with long-term effects. However, our primary efforts are concentrated on lower proportions (below 30\%), as this work is exploratory, and higher penetration scenarios may benefit from more advanced strategies such as platooning \citep{resultsPlatooning}.

\begin{figure}[th!]
    \centering
    \includegraphics[width=0.5\linewidth]{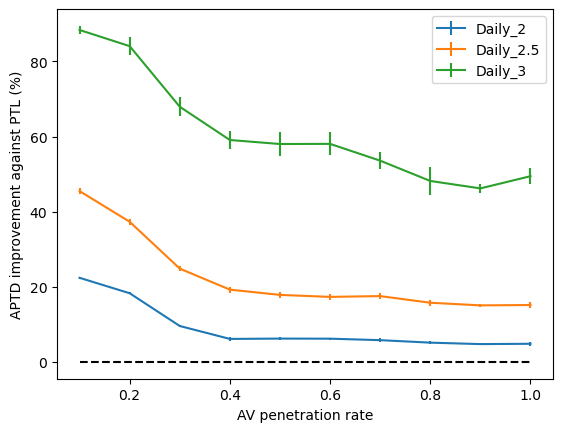}
    \caption{APD Change (in \%) of CAVDynamic\_24 Compared to DBL Across Different Demands and Vehicle Type Distributions.}
    \label{fig12:APD_baseline}
\end{figure}

Figure \ref{fig12:APD_baseline} highlights the above observations by focusing on a specific dynamic policy compared to the DBL baseline. The relative improvement is measured compared to the baseline DBL, and the error bars are calculated using the empirical standard deviation. It is evident that for lower proportions, there is a substantial improvement, whereas for higher proportions and certain demand levels, the improvement is small. This outcome is intuitive, as higher demand scenarios experience most of the delay outside the road network, specifically while waiting to enter the simulation. As a result, the scope for policy intervention is limited, leading to similar performance between the baseline and the proposed policy.

\subsection{Encouraging Transition to CAVs}
\begin{table*}[!ht]
\centering
\resizebox{\textwidth}{!}{%
\begin{tabular}{|c|c|l|l|l|l|l|l|l|l|l|l|l|}
\hline
\textbf{Demand}             & \textbf{vType}          & \multicolumn{1}{c|}{\textbf{numPass}} & \multicolumn{1}{c|}{\textbf{0}} & \multicolumn{1}{c|}{\textbf{10\%}} & \multicolumn{1}{c|}{\textbf{20\%}} & \multicolumn{1}{c|}{\textbf{30\%}} & \multicolumn{1}{c|}{\textbf{40\%}} & \multicolumn{1}{c|}{\textbf{50\%}} & \multicolumn{1}{c|}{\textbf{60\%}} & \multicolumn{1}{c|}{\textbf{70\%}} & \multicolumn{1}{c|}{\textbf{80\%}} & \multicolumn{1}{c|}{\textbf{90\%}} \\ \hline
\multirow{6}{*}{Daily\_2}   & \multicolumn{1}{l|}{HDV} &                                       & 192.8075                        & 174.4502                          & 178.586                           & 186.6696                          & 190.8419                          & 193.6343                          & 196.8692                          & 197.7708                          & 204.2912                          & 201.317                           \\ \cline{2-13} 
                            & \multirow{5}{*}{CAV}     & 1                                     &                                 & 131.0794                          & 135.8888                          & 148.8604                          & 161.7307                          & 170.5922                          & 177.5414                          & 183.7817                          & 188.0875                          & 192.2795                          \\  
                            &                         & 2                                     &                                 & 123.2912                          & 119.0482                          & 125.8434                          & 137.7245                          & 145.6153                          & 153.4509                          & 158.6836                          & 163.5224                          & 165.8215                          \\  
                            &                         & 3                                     &                                 & 121.9301                          & 108.3618                          & 109.0211                          & 117.0173                          & 123.8617                          & 131.4685                          & 136.6799                          & 140.2904                          & 142.6441                          \\  
                            &                         & 4                                     &                                 & 119.5082                          & 105.7319                          & 101.3724                          & 104.969                           & 110.9661                          & 112.0425                          & 117.3363                          & 118.5929                          & 119.6316                          \\  
                            &                         & 5                                     &                                 & 118.8228                          & 105.9128                          & 100.6134                          & 99.32158                          & 102.9467                          & 106.3301                          & 108.1566                          & 109.3695                          & 109.5974                          \\ \hline \hline
\multirow{6}{*}{Daily\_2.5} & \multicolumn{1}{l|}{HDV} &                                       & 193.2504                        & 181.7295                          & 192.1111                          & 204.0548                          & 209.8675                          & 212.1832                          & 216.5011                          & 218.646                           & 223.8874                          & 220.4699                          \\ \cline{2-13} 
                            & \multirow{5}{*}{CAV}     & 1                                     &                                 & 131.6632                          & 143.3066                          & 160.0464                          & 174.2366                          & 185.389                           & 193.9482                          & 200.653                           & 207.0123                          & 210.806                           \\  
                            &                         & 2                                     &                                 & 123.1039                          & 121.3499                          & 129.8217                          & 142.777                           & 154.4216                          & 164.8314                          & 169.3807                          & 174.6785                          & 180.3309                          \\  
                            &                         & 3                                     &                                 & 120.8517                          & 110.141                           & 109.8906                          & 118.9824                          & 126.7184                          & 135.6798                          & 141.7187                          & 145.0163                          & 148.8438                          \\  
                            &                         & 4                                     &                                 & 122.0472                          & 108.9387                          & 102.2998                          & 107.4497                          & 109.7095                          & 116.0933                          & 119.4                             & 122.3172                          & 124.7952                          \\  
                            &                         & 5                                     &                                 & 121.4315                          & 110.2402                          & 101.7723                          & 102.688                           & 105.1889                          & 107.6653                          & 109.3261                          & 110.4058                          & 112.5789                          \\ \hline \hline
\multirow{6}{*}{Daily\_3}   & \multicolumn{1}{l|}{HDV} &                                       & 142.8618                        & 57.50357                          & 69.85652                          & 107.2592                          & 117.6958                          & 121.1946                          & 123.5185                          & 131.5189                          & 137.3365                          & 137.0107                          \\ \cline{2-13} 
                            & \multirow{5}{*}{CAV}     & 1                                     &                                 & 38.99002                          & 51.26042                          & 82.72547                          & 96.843                            & 105.1141                          & 110.5352                          & 119.4374                          & 124.5683                          & 131.8042                          \\  
                            &                         & 2                                     &                                 & 37.81332                          & 43.11235                          & 65.98568                          & 77.90601                          & 86.70729                          & 91.14111                          & 98.91568                          & 104.9222                          & 111.2531                          \\  
                            &                         & 3                                     &                                 & 35.8457                           & 42.34916                          & 57.16978                          & 64.7012                           & 68.94654                          & 73.11388                          & 79.75148                          & 82.56191                          & 89.15968                          \\  
                            &                         & 4                                     &                                 & 33.70021                          & 38.87541                          & 51.62696                          & 60.25454                          & 64.80327                          & 65.55316                          & 69.22968                          & 71.92686                          & 74.78586                          \\  
                            &                         & 5                                     &                                 & 35.92674                          & 37.21651                          & 49.47757                          & 57.46019                          & 58.10816                          & 59.49245                          & 63.52464                          & 64.90312                          & 70.35529                          \\ \hline
\end{tabular}%
}
\caption{timeLoss (Defined in Section \ref{eq:APD}) of CAVs and HDVs Across Different Scenarios and Proportions Using CAVDynamic\_24.}
\label{tab:CAV_HD_comparison}
\vspace{-3mm}
\end{table*}

We now evaluate whether the proposed hybrid traffic law achieves the second objective: encouraging passengers to switch to CAVs and carpool. Table~\ref{tab:CAV_HD_comparison} compares timeLoss (Equation \ref{eq:APD}) between CAVs with varying passenger counts and HDV vehicles under a dynamic policy. timeLoss is used instead of APD to focus on vehicle experiences within the network, as departDelay treats all vehicles equally without prioritization.

The dynamic policy occasionally improves HDV timeLoss but often results in only minor improvements or slight detriments. In contrast, CAVs consistently experience significantly lower timeLoss. At low proportions (e.g., 10\%), CAVs see approximately 30\% less timeLoss than HDVs, as the threshold is typically set to 1, allowing all CAVs access to the restricted lane. This provides a strong incentive for HDV passengers to switch to CAVs, as discussed in Section~\ref{goals}.

As the proportion rise, incentives to switch to CAVs grow stronger, and timeLoss differences based on passenger count become evident. At 30\% Proportion, CAVs with one passenger experience more delays than those with three or more passengers. Higher proportions further widen this gap, reinforcing the incentive to carpool. These trends, driven by self-interest, encourage a shift toward carpooling in CAVs.

Such results cannot be achieved with a static policy, as scenario-specific adjustments are required. The dynamic policy, however, adapts automatically to achieve these outcomes.

\subsection{Rationale for Restricted-Lane Deployment}
\begin{figure}[H]
    \centering
    \includegraphics[width=0.5\linewidth]{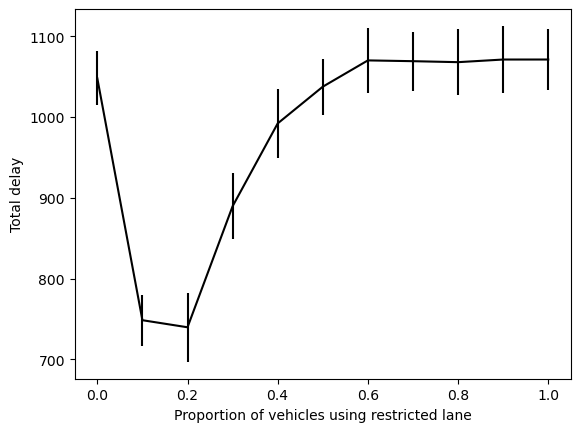}
    \caption{
The VD under demand scenario \textit{Constant\_3000} analyzed for varying proportions of vehicles permitted to access the restricted lane. The error bars are based on the standard deviation of the measured data.}
    \label{fig:totalDelay_motivation}
\end{figure}
\begin{figure*}[th!]
    \centering
    \includegraphics[width=\linewidth]{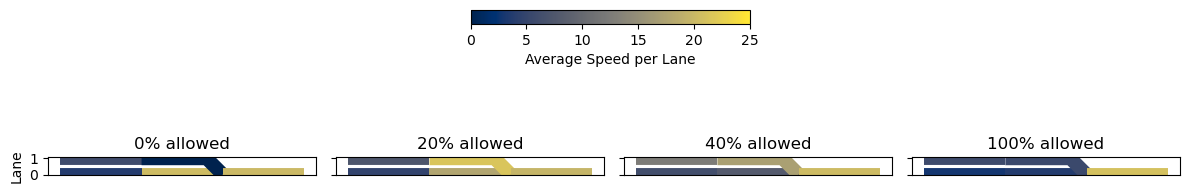}
    \caption{
The average speed across each segment of the road under demand scenario \textit{Constant\_3000}, analyzed for varying proportions of vehicles permitted to access the restricted lane.}
    \label{fig:AverageSpeed_motivation}
\end{figure*}
Figures \ref{fig:totalDelay_motivation} and \ref{fig:AverageSpeed_motivation} demonstrate that optimal traffic efficiency is achieved when approximately 20\% of vehicles are granted access to the restricted lane. Figure \ref{fig:AverageSpeed_motivation} shows that at this proportion, merging occurs smoothly, with minimal congestion. Conversely, allowing 40\% or more vehicles leads to inefficient merging and reduced speeds across the road. Figure \ref{fig:totalDelay_motivation} further confirms that the lowest vehicle delay (VD) occurs at the 20\% access point, with delays increasing significantly when access is unrestricted. These results align with the above outcomes, which indicate a low APD for policies that allow about 10-20\% of the vehicles to use the restricted lane.

\subsection{Application to a Real-World Network Scenario}
\begin{figure*}[th]
    \centering
        \centering
        \includegraphics[width=0.8\linewidth]{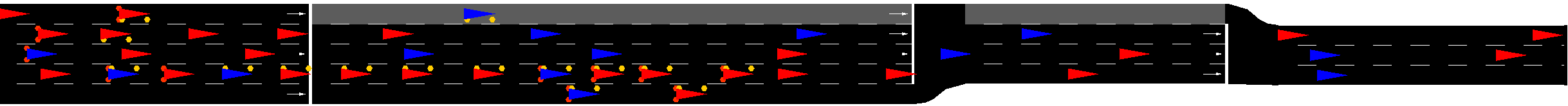}

    \caption{Illustration of the Configuration of the Case Study Road Network.}
        \label{fig:network_casestudy_overview}
\end{figure*}
\begin{table*}[ht]
\centering
    \vspace{2ex}
    % subfigures for Data Table:
    \begin{subtable}{\textwidth}
    \centering
    \resizebox{0.7\textwidth}{!}{%
    \begin{tabular}{lcccccc}
     & DBL & Plus\_1 & Plus\_2* & Plus\_3* & Plus\_4* & Plus\_5* \\
    \hline
    \textbf{APD} & \underline{1152.62} & 1565.54 & 1417.08 & \textit{1010.18} & 1084.55 & 1112.22 \\
    \end{tabular}%
    }
    \caption{Baseline Policies (only HDVs)}
    \label{tab:casestudy_baseline}
    \end{subtable}
    
\begin{subtable}{\textwidth}
  \centering
  \resizebox{\textwidth}{!}{%
\begin{tabular}{l|cccccccccc}
    \textbf{Policy / CAVs Proportion} & \textbf{10\%} & \textbf{20\%} & \textbf{30\%} & \textbf{40\%} & \textbf{50\%} & \textbf{60\%} & \textbf{70\%} & \textbf{80\%} & \textbf{90\%} & \textbf{100\%} \\
    \hline
        CAVStaticPlus\_1 & 1014.34 & 1087.72 & \textbf{1268.64} & \textbf{1460.58} & \textbf{1570.46} & \textbf{1705.98} & \textbf{1725.97} & \textbf{1638.25} & \textbf{1608.00} & \textbf{1565.54} \\
    CAVStaticPlus\_2 & 1060.07 & 1033.63 & 1014.22 & 1021.45 & 1078.66 & 1131.58 & \textbf{1205.05} & \textbf{1272.57} & \textbf{1333.96} & \textbf{1417.08} \\
    CAVStaticPlus\_3 & 1133.22 & 1104.56 & 1075.92 & 1081.11 & 1069.82 & 1045.00 & 1041.62 & 1026.44 & 1015.36 & 1010.18 \\
    CAVStaticPlus\_4 & 1130.82 & 1123.57 & 1132.54 & 1120.85 & 1117.66 & 1105.04 & 1101.02 & 1105.22 & 1077.16 & 1084.55 \\
    CAVStaticPlus\_5 & 1139.60 & 1142.40 & 1130.19 & 1139.06 & 1137.56 & 1124.12 & 1135.59 & 1125.72 & 1133.92 & 1112.22 \\
    CAVDynamic\_22 & 1016.48 & 1068.35 & 1146.18 & \textbf{1190.47} & \textbf{1203.44} & \textbf{1207.59} & \textbf{1200.24} & \textbf{1222.86} & \textbf{1222.68} & \textbf{1200.52} \\
    CAVDynamic\_23 & 1024.54 & 1050.41 & 1106.41 & \textbf{1155.43} & \textbf{1158.92} & 1148.86 & \textbf{1168.45} & \textbf{1159.49} & \textbf{1175.56} & \textbf{1171.54} \\
    CAVDynamic\_24 & 1026.60 & 1039.20 & 1084.42 & 1114.84 & 1106.55 & 1106.46 & 1126.43 & 1123.59 & 1131.54 & 1129.37 \\
    CAVDynamic\_25 & 1028.13 & 1054.29 & 1068.98 & 1098.85 & 1100.84 & 1093.60 & 1084.32 & 1100.31 & 1107.32 & 1112.56 \\
\end{tabular}
\vspace{2ex}
}
    \caption{Dynamic and Static Policies}
    \label{tab:dyanmic_casestudy}
    \end{subtable}
\caption{APD results for the case study data. Bold, italic, and underlined values as well as * follow the same conventions as in Table \ref{tab12:APD}.}
\label{tab:casestudy_APD}
\end{table*}

\begin{figure}[t]
    \centering
        \centering
        \includegraphics[width=0.7\linewidth]{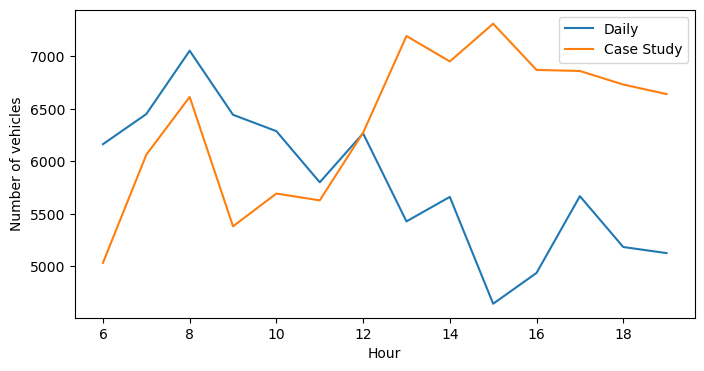}
    \caption{Demand per Hour for \textit{Daily\_k} and the Case Study.}
        \label{fig:DemandFig}
\end{figure}

To further validate the findings, a more realistic case study was conducted, as illustrated in Figure~\ref{fig:network_casestudy_overview}. In this scenario, a five-lane roadway narrows to three lanes: the rightmost lane is eliminated first, followed by the left lane, which currently serves as a DBL. The study segment is located on Ayalon Road (northbound), near its junction with Road~5. A dedicated bus lane (DBL) is currently operating upstream, but to accommodate the proposed dynamic policies, this initial segment was modeled with five general-purpose lanes. As presented in Figure \ref{fig:DemandFig}, minor adjustments were made to the demand profile to account for a shifted measurement point (compared to Section~\ref{def:Daily}). However, overall volumes remain comparable, apart from a heightened evening peak.

Table~\ref{tab:casestudy_APD} summarizes the Average Passenger Delay (APD) across the policies for this realistic scenario. While \textit{CAVDynamic\_24} and \textit{CAVDynamic\_25} consistently outperform the DBL baseline, their advantages are less pronounced than those reported in Table~\ref{tab12:APD}. Additionally, certain dynamic policies (e.g., \textit{CAVDynamic\_22}) do not invariably surpass DBL performance, suggesting that even though these strategies can enhance traffic flow, their applicability to more complex networks is far from straightforward. Achieving robust outcomes in such contexts likely requires more sophisticated approaches that rely on broader traffic data rather than a single local measurement of the restricted lane.

Further experiments that include the intersection with Road~5 demonstrate suboptimal performance across all policies. In these instances, \textit{Plus\_1}, effectively eliminating the restricted lane, produces the best results. This outcome underscores the challenges of implementing hybrid traffic laws at scale or under highly congested conditions, reinforcing the need for additional research into advanced dynamic control strategies that account for larger and more intricate network configurations.

\section{Conclusions}
This research presents a novel approach to traffic management by introducing hybrid traffic laws tailored to mixed environments of autonomous and human-driven vehicles. The proposed policies leverage the unique capabilities of connected autonomous vehicles to enforce lane restrictions and optimize traffic flow. Simulation experiments found that dynamic policies, which adjust access thresholds in real-time, consistently outperform baseline policies in reducing average passenger delay (APD). These improvements are most notable at low proportions of connected autonomous vehicles, where dynamic policies incentivize the incorporation of CAVs and promote shared mobility practices, such as carpooling.

In a more complex, realistic corridor study, dynamic strategies continued to yield better performance than baseline scenarios, although the magnitude of improvement was reduced. This finding underscores the challenges of translating adaptive policies into real-world settings, where additional bottlenecks—such as merges and off-ramps—complicate traffic patterns. As CAV proportions increase, certain static policies can deliver comparable performance, particularly in high-demand situations, indicating the need for adaptive regulations that can operate effectively across diverse traffic conditions.

By granting restricted lane access exclusively to CAVs with higher occupancy rates, the policies create clear advantages for travelers who opt for shared mobility solutions. This not only accelerates the transition to autonomous technology but also encourages more efficient travel patterns, reducing overall vehicle congestion and environmental impacts. However, as the proportion of CAVs increases, static policies can achieve comparable performance, particularly under high-demand conditions. This highlights the importance of adaptable policies to accommodate diverse traffic scenarios. Our findings contribute valuable insights into the design of enforceable and effective traffic laws, promoting a smoother transition towards a more sustainable and efficient transportation system.

Future work could investigate dynamic policies capable of maintaining their advantages in more complex network configurations, including those with multiple merges and congestion hotspots. In addition, other hybrid traffic laws merit exploration—such as variable speed limits informed by CAV data, more sophisticated merging controls on highway ramps, or automated clearance for emergency vehicles.

\section*{Acknowledgements}

This research was jointly funded by the Ministry of Innovation, Science and Technology, the Ministry of Transport, the National Road Safety Authority, and Netivei Israel.
\clearpage
\bibliographystyle{named}
\bibliography{references}

\end{document}